\documentclass[11pt]{article}                  
  \usepackage{times}                   

 \usepackage[dvips]{graphicx}

\begin{document}

\title{Magnetic turbulence in the plasma sheet}
\author{Z. V\"{o}r\"{o}s (1), 
W. Baumjohann (1),
R. Nakamura (1),  M. Volwerk (1),\\ 
A. Runov (1), T. L. Zhang (1), H. U. Eichelberger (1), \\
R. Treumann (2), E. Georgescu (2), A. Balogh (3), \\
B. Klecker (2) and H. R\`{e}me (4) \\
(1) Institut f\"{u}r Weltraumforschung der \"{O}AW, Graz, Austria,\\
(2) Max-Planck-Institut f\"ur extraterrestrische Physik, Garching, Germany, \\
(3) Imperial College, London, UK, \\
(4) CESR/CNRS, Toulouse, France. 
}
\date{}
\maketitle

\begin{abstract}
Small-scale magnetic turbulence observed by the Cluster spacecraft in the plasma sheet is
investigated by means of a wavelet estimator suitable for detecting distinct scaling
characteristics even in noisy measurements. 
The  spectral estimators  used for this purpose are affected by a frequency dependent bias.
The variances of the wavelet coefficients, however, match the power-law shaped spectra, which makes
the wavelet estimator essentially unbiased.
These scaling characteristics 
of the magnetic field data
appear to be essentially 
non-steady and intermittent. The scaling properties of bursty bulk flow (BBF)  
and non-BBF  associated magnetic fluctuations are analysed with the aim of understanding  processes
of energy transfer between scales. Small-scale ($\sim 0.08-0.3$ s) 
magnetic fluctuations having the same scaling index $\alpha \sim 2.6$ as the large-scale 
($\sim 0.7-5$ s) magnetic fluctuations occur during BBF-associated periods.
During non-BBF associated periods the energy transfer to small scales is absent, and the large-scale
scaling index $\alpha \sim 1.7$ is closer to Kraichnan or Iroshnikov-Kraichnan scalings.
The anisotropy characteristics of magnetic fluctuations show both scale-dependent and scale-independent
behavior. The former can be partly explained in terms of the Goldreich-Sridhar model of MHD turbulence, which
leads to the picture of Alfv\'{e}nic turbulence  parallel and of eddy turbulence  perpendicular
to the mean magnetic field direction. 
Nonetheless, other physical mechanisms, such as transverse magnetic structures, velocity shears, or 
boundary effects can contribute to the anisotropy characteristics of plasma sheet turbulence.
The scale-independent features are  related to  anisotropy characteristics which occur
during a period of magnetic reconnection and fast tailward flow.
\end{abstract}

%
%

%


\section{Introduction}
Magnetic field measurements on board  the Geotail and AMPTE/IRM spacecraft revealed that the power-law
behaviour of the energy density spectra in frequency space, $P(f) \sim f^{- \alpha}$  depends 
on the considered frequency range, both in the distant  tail [{\it Hoshino et al.}, 1994], 
and in the
near-Earth plasma sheet [{\it Bauer et al.}, 1995]. This finding was confirmed 
by {\it Ohtani et al.} [1995] using
fractal analysis of the fluctuations in the AMPTE/CCE and 
SCATHA magnetic field measurements. The spectral index changes around
$f_{c1} \sim 0.01 - 0.08$ Hz. 
The observed scaling indices are $\alpha_1 \sim 0.5 - 1.5$ for $f < f_{c1}$ and $\alpha_2 \sim 1.7 - 2.9$
when $f > f_{c1}$ [{\it Milovanov et al.}, 2001; {\it Volwerk et al.}, 2003, 2004]. 
The uncertainty of the estimated values of  $\alpha_1$ and $\alpha_2$ can be explained 
by slightly different frequency ranges studied by various authors and by the occurrence of spatially 
inhomogeneous, intermittent and transitory mechanisms involved in the generation of the observed scalings.

Since the characteristic timescale $\sim 1/f_{c1}$ was found to be a few times  the proton gyroperiod, ions 
can play an important role in the excitation of the corresponding magnetic fluctuations. The high-frequency
part of the spectrum can be produced by an MHD cascading process. Instability mechanisms or 
self-organisation processes, such as coalescence of magnetic 
islands or inverse cascades were proposed for the explanation of low-frequency 
scaling [{\it Hoshino et al.}, 1994].

For frequencies higher than 0.1 Hz a second break in spectral power  density  
was predicted theoretically at a turnover frequency $f_{c2}$ [{\it Milovanov et al.}, 2001].
The estimation of the scaling index $\alpha_3$ for frequencies $f>f_{c2}$,
however, is encumbered by  noise produced by the magnetometer itself. 
In this paper we examine those high frequency fluctuations for which the spectral indices 
$\alpha_2$ and $\alpha_3$ are expected. To this end we use a wavelet method which allows to estimate
the scaling propeties of magnetic fluctuations over the mentioned frequency ranges.
Reliable spectral information could not be extracted, hitherto, from the noisy 
high-frequency magnetometer data. Nonetheless, in Section 2 we argue that an essentially
unbiased wavelet method combined with an embedding technique of synthetic scaling signals allows
us to estimate the scaling parameters reliably.

Very little is known about the origin of the multi-scale turbulence in the plasma sheet.
Its appearance is obviously 
dependent on the driving and dissipation mechanisms which characterize an observed fluctuating process
[{\it Borovsky and Funsten}, 2003].
Magnetic field and plasma flow fluctuations were observed by a number of authors 
[e.g., {\it Coroniti et al.}, 1980;  {\it Borovsky et al.}, 1997; 
{\it Neagu et al.}, 2002]. Plasma shear flows might appear as possible drivers of those fluctuations.
The statistical analyses of  AMPTE/IRM data in the near-Earth
magnetotail have already shown that there exist rapid plasma flows in the central plasma sheet 
which are preferentially perpendicular to the magnetic field and strongly sunward
oriented  [{\it Baumjohann et al.}, 1989, 1990]. 
Subsequent research indicated that such flows are preferentially organized into
sporadically occurring groups of bursty bulk flows (BBF), typically lasting for $~10$ min.  
Despite  their short duration, BBFs are the carriers of decisive amounts of mass,
momentum and magnetic flux [{\it Angelopoulos et al.}, 1992; {\it Angelopoulos et al.}, 1993;
{\it Sch\"{o}del et al.}, 2001]. 
Therefore BBFs can stir the plasma sheet plasma very efficiently. 
In this paper we place emphasis on comparison of multi-scale power and scaling properties of the magnetic fluctuations 
during BBF and non-flow or post-flow periods. We will also try to identify the
turnover frequency $f_{c2}$, comparing the estimated values of $\alpha_2$ and $\alpha_3$ during 
BBF and non-BBF periods.

The observed scaling features, together with those expected from  MHD considerations, 
will help us to classify the magnetic fluctuations observed by the Cluster spacecraft in the plasma sheet.
Recent approaches to MHD turbulence highlight the influence of  a mean magnetic field on scale-dependent
appearance of anisotropic fluctuations [{\it Goldreich and Shridhar}, 1995]. 
The anisotropy increases towards small scales. Therefore, the high-resolution Cluster magnetometry provides
a way for studying  the  scale-dependent anisotropy  of magnetic fluctuations. We discuss the observed magnetic 
anisotropies in terms of recent MHD theories and simulations.

\section{The wavelet estimator}
We analyse the scaling properties of 67 Hz resolution magnetic field fluctuations 
from the Cluster fluxgate magnetometer  [{\it Balogh et al.}, 2001] using a wavelet method proposed
by {\it Abry et al.,} [2000]. Wavelets offer an effective way for getting an unbiased estimate of $\alpha$ 
[{\it Veitch and Abry}, 1999], 
whereas power spectral estimations are affected by frequency-dependent multiplicative bias [{\it Abry et al.}, 1995]. 
The spectral estimation is biased mainly because it is performed over a constant bandwidth which
does not match the $f^{-\alpha}$ structure of the spectrum properly. The variance of discrete wavelet coefficients
(see later), however, matches well the power-law shaped spectrum. 
Unlike Fourier methods, wavelet methods work better for processes exhibiting spectral scaling 
because wavelets have  inbuilt rescaling properties.
The wavelet estimator allows to 
identify different kinds of scaling processes which are invisible to biased spectral estimators, especially
in noisy and short data.

\subsection{Outline of the method}
{\it Abry et al.,} [2000] proposed a semi-parametric wavelet technique based on a fast pyramidal filter bank 
algorithm for the estimation of scaling parameters $c_f$ and $\alpha$ in the relation $P(f)\sim c_f f^{-\alpha}$,
where $c_f$ is a nonzero constant. The algorithm consists of several steps. First, a discrete wavelet 
transform of the data is performed over a dyadic grid $(scale, time)=(2^j,2^j t)$ and $j,t 
\in \bf{N}$. Then, at each octave $j=log_2 2^j$, the variance $\mu_j$ of the discrete wavelet coefficients 
$d_x(j,t)$  is computed through
\begin{equation}
\mu_j = \frac{1}{n_j}\sum_{t=1}^{n_j}d_x^2(j,t) = 2^{j\alpha}\ c_f\ C(\alpha, \psi_0)
\end{equation}
where $n_j$ is the number of coefficients at octave $j$, and
$ C(\alpha, \psi_0)=\int f^{-\alpha} \mid \psi_0(f)\mid^2 df$, $\psi_0(f)$ stands for the power spectrum 
of the mother wavelet $\Psi_0$. Finally, from Equation (1) $\alpha$ and $c_f$ are estimated by constructing 
a plot of $y_j\equiv log_2 \mu_j$ versus $j$ (logscale diagram, LD) and by using a weighted linear regression 
over the region  $(j_{min}, j_{max})$ where $y_j$ is a straight line. In this paper we use the Daubechies
wavelets for which finite data size effects are minimized and the number of vanishing moments $M$ can be
changed. The latter allows to cancel or decrease the effects of linear or polynomial trends and ensures that
the wavelet details are well defined.

We note that the wavelet estimator, Equation (1), is also affected by a multiplicative bias (multiplication
by $C(\alpha, \psi_0)$, see later), but $C$ is not frequency dependent. A similar multiplicative term
for spectral estimators is frequency dependent, which makes the log-log plot of the power versus frequency
not precisely linear. Therefore the estimation of $\alpha$ via linear regression in the log-log plot
is strongly affected by bias [{\it Abry et al.,} 1995]. Within the wavelet approach {\it Veitch and Abry} [1999]
have found explicit solutions for the variances of $log_2(\mu_j)$ which are then used  in the weighted
regression in the LD. These improvements make the wavelet estimator unbiased even for data of small length.

Different kinds of scaling can be analysed by this technique. The value of $\alpha$ and the scaling range
$(j_m,j_n)$ are characteristic for the specific scalings present [{\it Abry et al.,} 2000]. For example 
$\alpha \in (0,1)$ at 
the largest scales present in the data is characteristic for long range dependent processes, while
$\alpha \sim 0$ at large scales together with $\alpha \in (0,1)$ at small scales is a signature of noise.
Furthermore, $\alpha \in (1,3)$ aligned at almost all scales available corresponds to a nonstationary, 
self-similar (fractal) process having a fractal dimension $D=(5-\alpha )/2$ [{\it Abry et al.,} 2000].

\subsection{An embedding technique}
When the scaling index is estimated over a high frequency range, magnetometer noise can strongly influence
the outcome. Since different estimation techniques used over  slightly different frequency ranges, 
there is a large scatter in the previously observed values of $\alpha$, which makes it difficult
to compare the scaling properties in turbulence.
Furthermore, the estimated values are presumably affected by 
the occurrence of transitory processes in the plasma sheet. We therefore 
perform the estimation of $c_f$ and $\alpha$ within 
sliding overlapping windows of width $W$  with a time shift $S<<W$ to analyse the data over different frequency 
ranges with the same method and simultaneously take into account their temporal evolution. 

In order to examine how the signal-to-noise ratio (SNR, the relative power of the signal and the noise in LD)  
influences the estimation of 
$\alpha $ within the
window $W$ in small-scale fluctuations,  we generate a self-similar fractal signal with known second order scaling properties
using wavelets  and embed it in a time series of quiet time magnetic field measurements from
the Cluster spacecraft. As a synthetic fractal signal we use fractional Brownian motion (fBm) 
generated by a fast filter bank based pyramidal algorithm [{\it Abry and Sellan,} 1996]. 
Fractional Brownian motion has a Gaussian distribution of increments by definition. 
At the same time, magnetic fluctuations
exhibit two-point Gaussian statistics only for large separations. Therefore, we consider the 
fBm process as an approximation which can roughly model the second order properties of the 
magnetic fluctuations.

We suppose that the magnetic fluctuations observed in the plasma sheet 
during quiet times are close
to the magnetometer noise level. Figure 1a shows the magnetic field $B_X$ component (67 Hz resolution) observed 
by Cluster 3 on September 7, 2001. If not specified the GSM coordinate system
will be used throughout the paper.
Two short fractal signals $A$ and $B$, with the same scaling index 
$\alpha _A\equiv \alpha _B = 1.6$ but different amplitude are added  to $B_X$. In Figure 1a $B_X$ is depicted from 
1840 to 1917 UT. First, signal A is added into $B_X$ from $\sim$1857 to $\sim$1901 UT. The average amplitude 
of fluctuations of A is much larger than of $B_X$.
In Figure 1b a dashed horizontal line show the theoretical value of $\alpha _A\equiv \alpha _B = 1.6$.
The estimated $\alpha _A$ reaches this theoretical value in a wider interval in Figure 1b as was the length of 
signal A in Figure 1a. Despite the non-precise timing,  the estimated value of $\alpha _A$ is correct.
Next, signal B is embedded into $B_X$. Signal B and $B_X$ have  comparable average amplitude.
The estimated $\alpha _B$ in Figure 1b does not achieve the theoretical
horizontal line, therefore $\alpha _B$ is underestimated. 
We can conclude that, whenever the amplitude of the signal is significantly larger than the amplitude 
of the noise (case A), the
correct value of $\alpha$ is recovered, otherwise $\alpha$ is underestimated (case B).

Let us explain now the reason of non-precise timing.
In Figure 1b $\alpha s$ are estimated within a sliding window $W\sim 480 $ s
over time scales (octaves) $j_1 = 2  $ and $j_2 = 4$ ($\sim 0.08-0.33$ s and 
$\sim 3-12$ Hz; these are approximate Fourier periods and
frequencies that correspond to the main oscillations within the Daubechies wavelets). 
The embedded fractal signals A and B 
produce a rapid departure of the estimated values $\alpha$ from the noise level $\alpha \sim 0.3$, as soon
as the analysing window includes some part of a fractal signal (A or B in Figure 1a). This effect is visible
at $\sim$1853 UT and before 1906 UT, when two  analysing windows depicted by horizontal lines in 
Figure 1b are centered in those time positions. The dashed, vertically extended right and left ends of the windows 
match perfectly the beginnings and the ends of the fractal signals A and B. When the timing is important, the effect
of window length has to be accounted. 

Another effect which influences the estimation of the scaling index $\alpha$. 
If the duration of a scaling process $T$ is less than  half the window length, $W/2$, $\alpha$ is also underestimated.
In Figure 1a the length of the fractal signals A and B  equals $W/2$.
The proper sliding window length for unbiased estimation of $\alpha$ can be found by  trial. For
decreasing $W$ the estimated $\alpha$ increases, then saturates. Obviously, as $W$ decreases, 
the fluctuations  of the estimations increase.  In general, $W$ has to be chosen such that the opposing 
requirements for robustness of scaling parameter estimations (wide window needed)  and for time 
localization of short events (narrow window needed) were matched. For the events analysed in this paper
$W = 61$ s was found to be an optimal choice.

Finally, let us consider possible kinds of non-physical mechanisms which can lead to spurious scalings. Time
series analysis methods may lead to spurious results when applied to short data series containing
irregularities, e.g. jumps, spikes [{\it Katsev and L'Heureux,} 2003] or random pulse trains 
[{\it Watkins et al.,} 2001]. Spikes, i.e., single values with large deviation are removed during quality
check of the data. Possible effects of  other irregularities will be discussed jointly with the
estimation of the scaling parameters in subsequent sections.

\subsection{Noise correction}
The above explained embedding technique allows us to introduce corrections,  whenever the amplitude of a fractal
signal approaches the noise level. It can happen when high frequency (small-scale) scalings are examined.
A few more steps are needed to explain the method in detail. To this end
we  show
the  typical scaling properties
of a fractal signal ($s$), noise ($n$) and of 
a compound additive
process (signal+noise, $sn$)  in LD. For what follows the subscripts $s, n, sn$ will correspond to 
the  specified processes. We also demonstrate that a single correction curve (see later) 
belongs to every embedded model fractal signal with a given $\alpha_s$. 

Equation 1 describes the scaling properties of a signal through the scaling of  the corresponding wavelet 
coefficients. Taking a base 2 logarithm of both sides of Equation 1, we obtain an equation
of  line in LD. Therefore any scaling signal (a straight line in LD)
can be characterized by a pair of
values $(c_f(j), \alpha(j_{min}, j_{max}))$, where $c_f(j)$ is the power of intercept at  octave $j$.
Here we suppose that, for a fixed mother wavelet, the last term, $C(\alpha, \psi_0)$, on the right hand side of
Equation 1, is dynamically unimportant. The change of $\psi_0$, however, can  influence 
the estimation of scaling parameters. We use the Daubechies family of wavelets with
changing number of vanishing moments $M$. It does slightly influence our estimations and the corresponding
uncertainities represent a source of error. 

Figure 2 shows  an exemple of LD.
The magnetometer noise is depicted by a single curve, since we suppose, it is time-independent (see later).
It exhibits signatures typical for noise: $\alpha_n \sim 0$ at large scales ($j>7$) together 
with $\alpha_n \sim 0.3 \in (0,1)$ at small scales ($j<6$) [{\it Abry et al.} 2000].
For the sake of clarity the straight lines corresponding to the scalings of both signal and  compound process 
are depicted only at the left hand side of Figure 2 ($j\le5$). 
The amplitude of the model fractal signal was changed from values larger to values 
smaller than the amplitude of the noise. The resulting   parallel, shifted straight lines are 
representations of self-similar signals in LD. The compound process exhibits similar characteristics
as the signal itself,  while  SNR is large. If SNR becomes $\le 1$, the compound process approaches
the straight line of noise at the scales $j_1$ and $j_2$. 

Time-independent noise has  averages $(\langle c_{fn}(j)\rangle, \langle \alpha_n(j_1, j_2)\rangle) \sim const.$ 
Our goal is to estimate the time evolution of the scaling exponent of the signal $\alpha_s(j_1, j_2, t)$
for known $(c_{fn}, \alpha_n)$ and $(c^{'}_{fsn}(j,t), \alpha_{sn}(j_1, j_2, t))$.
The normalization $c^{'}_{fsn}(j,t) \equiv c_{fsn}(j, t) / \langle c_{fn}(j)\rangle$ 
ensures that for weak signals,
$c_{fsn} \sim c_{fn}$ (noise dominates), and   $c^{'}_{fsn} \sim 1$. Actually, $c^{'}_{fsn}(t)$
represents a time dependent SNR.

Now we can use the introduced notation and the embedding technique   to study
the dependence of $\alpha_{sn}$ on $c^{'}_{fsn}$ within a window $W$. Figure 3 shows the relationship
$\alpha_{sn}(c^{'}_{fsn})$ (correction curves),  which was obtained experimentally by embedding
two model fractal signals with $\alpha_{mod} = 1.5$ and $2$, gradually decreasing their amplitudes
(decreasing SNR), into quiet time $B_X$ measured by s/c 4 on  September 7, 2001. For both model fractals
the correction curves were
computed at the
octaves $j=0$ and $j=4$, separately. Independent of $j$, $\alpha_{sn}$ approaches asymptotically
$\alpha_s \equiv \alpha_{mod}$ for large enough $c^{'}_{fsn}$. Large $c^{'}_{fsn}$ means that 
SNR is large and the compound signal exhibits the same scaling as the fractal model signal 
(left-top straight lines in Figure 2).
If $c^{'}_{fsn} \sim 1$, the noise
dominates, and $\alpha_{sn} \sim \alpha_{n}$ (scaling of the compound process between $y_j=-5$
and $-10$ in Figure 2). 
Between the two limits ($c^{'}_{fsn} \to \infty$ and $c^{'}_{fsn} \sim 1$), however, the course of the curves
for $j=0$ and $j=4$ in Figure 3 is rather different. 
Because of the complex behaviour of the compound process at $j=0$ near the noise level
(Figure 2), $\alpha_{sn}$ cannot be determined from the $\alpha_{sn}(c^{'}_{fsn})$ relationship
unequivocally (Figure 3). Since the relation is unambiguous at the scale $j=4$, we will consider
the $\alpha_{sn}(c^{'}_{fsn})$ curves at that octave throughout the paper. 

The accuracy of the wavelet method estimation of $\alpha$ can be controlled by the power of intercept
of the compound process $c^{'}_{fsn}$. Figure 3 shows that for $c^{'}_{fsn}<<100$ we find 
$\alpha_{sn} < \alpha_{mod}$, therefore $\alpha_s$ is underestimated. The $\alpha_{sn}(c^{'}_{fsn})$
relationship allows us to introduce corrections and estimate the effective  asymptotic value of $\alpha_s$. 

We note that $c_{fsn}(t)$ and $\alpha_{sn}(j_1,j_2,t)$ are computed directly from magnetic field time
series by using a sliding window W. The key assumption of the method is time-independence of the scaling 
parameters for noise. Obviously, this condition is not strictly fulfilled. Therefore, it is important
to examine the time-independence of $\langle c_{fn}\rangle$ and $\langle \alpha_{n}\rangle$. 
The parameters $c^{'}_{fsn}(t)$ and $\alpha_{sn}(j_1,j_2,t)$ computed from 
magnetic field $B_X$ component (67 Hz resolution) observed by the Cluster spacecraft s/c 1 and 3 during the
quiet interval 1700-2100 UT on September 7, 2001, are depicted in Figure 4. Here $W = 68$ s and $S = 4$ s.
During the chosen interval the spacecraft are traversing the neutral sheet from the northern ($B_X \sim 18$ nT) to 
the southern hemisphere ($B_X \sim -15$ nT) (Figure 4a) and the plasma flow velocity does not
exceed 50 km/s (not shown). Independent of the location of the Cluster spacecraft and despite the
occurence of low frequency wave motion after 1830 UT, both  $\alpha_{sn}(j_1,j_2,t)$ and $c^{'}_{fsn}(t)$ 
exhibit stationary behaviour, fluctuating around the mean values (Figure 4 b,c). 
The estimated $\alpha_{sn}$ can be
related to the magnetometer noise $\alpha_{n}$ because $\alpha_{sn}\sim \alpha_{n} \in (0,1)$
and $c^{'}_{fsn} \sim c_{fn} \sim 1$ (see also Figure 3). Noteworthy is that $\alpha_n$ and
$c_{fn}$ are slightly different for each s/c (Figure 4b,c) and also vary with the magnetic field
components. The examination of several quiet time magnetic field time series and time series
obtained from ground-based calibration measurements shows that in average (including the four s/c
and each magnetic component) $\langle \alpha_{n}\rangle = 0.5 \pm 0.25$ and 
$\langle c_{fn}\rangle = 0.004 \pm 0.003$. Similar 
values were obtained using 128 Hz quiet time magnetic field measurements from 
the Equator-S spacecraft (not shown).
For discrimination between $\alpha_s$ (which is expected to be $> 1$) and $\alpha_n$, this accuracy seems to be 
sufficient. It is expedient in concrete case studies, however, to compute the correction curves 
(Figure 3) for each spacecraft and component separately. 

Examining  $\alpha_{sn}$ and $c^{'}_{fsn}$ estimated for  $B_X$ from s/c 2 on September 7, 2001, 
we found short, but significant deviations from $\langle \alpha_{n}\rangle$ and $\langle c_{fn}\rangle$, 
respectively (not shown). The deviations occurred due to a $~ 5$ min long data gap on s/c 2, which led
to spurious scalings because of the step function-like irregularity in $B_X$, in accordance
with the findings of {\it Katsev and L'Heureux} [2003]. This type of spurious estimations can easily
be avoided by interpolation. Another type of jump-related spurious estimation might occur
near the plasma sheet boundary layer (PSBL) or owing to its motion. The PSBL 
cannot be identified from the magnetic field measurements alone[{\it Eastman et al.}, 1984]. 
Therefore, non-physical
scalings due to magnetic jumps near PSBL are not expected. 

In summary, using the wavelet approach, we can estimate directly 
from a time series the time evolution of the parameters of a compound process $(c^{'}_{fsn}(t), \alpha_{sn}(t))$.
The time independent parameters of noise $(\langle c_{fn}\rangle, \langle \alpha_n\rangle)$ 
can be obtained from quiet time measurements. When $c^{'}_{fsn}(t)$ is small, the effective asymptotic value of 
$\alpha_{s}(t)$ for a self-similar signal can be obtained from a correction curve taking a limit
$c^{'}_{fsn}(t)\to\infty$. Spurious estimations during quiescent periods can be avoided by careful
quality checks of the magnetic data.

\section{Magnetic fluctuations in the plasma sheet}
In order to introduce the wavelet method and the small-scale corrections to the estimated $\alpha$,
quiet time magnetic measurements together with artificial fractal signals were used in the previous
section. Since our goal is to investigate BBF and non-BBF related magnetic turbulence, here we
choose a $\sim4$h interval during which the Cluster spacecraft observed BBF and non-BBF intervals successively. 

\subsection{Event overview and small-scale corrections}
In what follows, we analyse burst mode (67 Hz) magnetic data from the Cluster fluxgate magnetometer (FGM)
[{\it Balogh et al.}, 2001] during the interval 0110-0500 UT on August 27, 2001, 
when the Cluster spacecraft were near the
$Z_{GSM}=0$ plane, in the postmidnight  magnetotail ($X_{GSM}\sim -19 R_E$).
The magnetic data are compared with the spin-resolution (~4 s) velocity data from the Cluster
ion spectrometry (CIS/CODIF) experiment [{\it R\'{e}me et al.}, 2001].
Figure 5a shows the $B_X$ component from s/c 1,3. The variation of the scaling index for two different
ranges of scales is depicted in Figure 5b. 
In addition to 
$(j_1, j_2)=(2, 4)$, corresponding time scale $\sim 0.08-0.33$ s), we also consider the octaves: 
$(j_3, j_4)= (5, 8) $, corresponding time scale  $\sim 0.7-5$ s .
Hereinafter we will refer to those scales as small scales $(j_1, j_2)$ and large scales 
$(j_3, j_4)$, respectively. Obviously, $j_1$ is limited by the resolution and $j_4$ by the chosen
window $W$.
The average noise levels $\langle \alpha_n\rangle$ for s/c 1,3 and $B_X$ magnetic components are also indicated by
black and gray dashed lines in Figure 5b.
The large-scale scaling exponents are anticorrelated with the magnitude of $B_X$ simply because of the
preferential occurrence of perpendicular flows closer to the neutral sheet ($B_X\to 0$ nT).
Figure 5c shows the variation of the normalized power of the intercepts $c^{'}_{fsn}$
computed for the  small scales $(j_1,j_2)$. Its time evolution exhibits periods of $c^{'}_{fsn} \sim 1$
intermittently interrupted by bursts of activity when $c^{'}_{fsn} > 1$. The former corresponds to the 
noise level, indicating that there is no significant energy flow from large scales into  small scales. 
The increased values of $c^{'}_{fsn}$ are associated with high frequency (small-scale) 
fluctuations of $B_X$ (Figure 5a), increased $\alpha_{sn}$ (relative to $\langle \alpha_{n}\rangle$, 
Figure 5b) 
and are well correlated with the appearance of bursty perpendicular flows (Figure 5d). 
The large scales  $(j_3,j_4)$ are 
not corrupted significantly by magnetometer noise. 

The effect of spurious estimations due to spikes, random pulses, or other types of nonstationary signals
can be addressed for BBF-associated periods, too. The use of Daubechies analysing wavelets with 
vanishing moments ensures that part of the nonstationarities represented by polynomial trends are 
cancelled. Spurious estimates of the scaling index might appear because of discontinuities
which separate earthward moving plasma structures from the background plasma ahead of them 
[{\it Sergeev et al.}, 1996]. In fact, before the appearance of the rapid perpendicular flow,
e.g., at ~0131 UT (Figure 5d), there is a large jump in the large scale scaling index from
$\alpha_{sn}(j_3,j_4)\sim 1$  to $2.5$.  This jump and all similar jumps in our data can partly be explained 
by the presence of a discontinuity in front of the plasma flow [{\it Nakamura et al.}, 2002]. 
On the other side, this short effect cannot be
separated from the non-physical effect of the sliding window, explained above (Figure 1).

After 0130 UT, $\alpha_{sn}(j_3,j_4)$ fluctuates around $\sim 2.5$ for more than 30 min (Figure 5b).
We believe, that all estimations exhibiting small fluctuations around a mean value for equal or longer 
time than the sliding window length ($\sim 1$ min), belong to a BBF-associated scaling process in our case.
Therefore these estimations are not spurious. The same is true for non-BBF associated processes. 
Also, $\alpha_{sn}$ is largely coherent between
s/c1 and 3. Short living pulses are supposed to be random, however. Moreover, some results of
multifractal approach  [{\it V{\"o}r{\"o}s et al.}, 2003] and of wavelet based structure function
approach [{\it V{\"o}r{\"o}s et al.}, 2004] also support the idea that BBF-associated magnetic
fluctuations correspond to a physical scaling process. 

During the BBF-associated periods 
one finds that $1<c^{'}_{fsn}<100$, therefore $\alpha_{s}$ is underestimated 
(see Figure 3). 
Since self-similarity is defined for $1 < \alpha < 3$, we used 4 model fractal time series with
$\alpha_{mod} = 1.5, 2, 2.5, 3$ and quiet time $B_X$ magnetic data from s/c 1,3 to compute the
$\alpha_{sn}(c^{'}_{fsn})$ correction curves at the octave j=4 (Figure 6). These curves can be used to 
introduce corrections and to estimate the effective value of $\alpha_s$. Like in Figure 3 there are
two limiting cases. The condition $c^{'}_{fsn} \to \infty$ leads to $\alpha_{sn}=\alpha_{s}=\alpha_{mod}$,
with slower approach to a model value for $\alpha_{sn}>2$. The limit $c^{'}_{fsn} \to 1$ 
leads to $\alpha_{sn}=\langle \alpha_{n}\rangle$. 
The error bars depicted at the left-top corner of Figure 6 
arise from 
the uncertainty of the estimation of $\alpha_{sn}$ in a logscale diagram using Daubechies wavelets with 
varying number of vanishing moments ($M=3-7$) and from the uncertainty of the  magnetometer 
noise level estimation within the chosen window W. 
The depicted error bars represent only the uncertainity of the determination
of the correction curves, and not the uncertainity of the estimations of $\alpha_{s}(j_1,j_2)$.
The estimation error of the small scale  $\alpha$  changes from $\pm 0.2$ to $\pm 0.5$ as
$c^{'}_{fsn} \to 1$. The errors can be estimated by adding more correction curves to Figure 6 (not shown).
It is more difficult to distinguish an $\alpha$ as $c^{'}_{fsn}$ decreases.
We present two examples how to read the measured $\alpha_{sn}(j_1,j_2)$
using both Figures 5 and 6. The two dashed lines A and B in Figure 6 correspond to two pairs of values
$(c^{'}_{fsn}, \alpha_{sn})$ in Figure 5b,c before 0330 UT. 
The gray lines and the letter A belong to s/c 3, while the black lines and the letter B belong to s/c 1.
In both examples the corrected effective 
value of small-scale $\alpha_{s}(j_1,j_2)$ is  close to the correction curves which
asymptotically reach the horizonthal line $\alpha_{mod}=3$ in Figure 6. 
After correction, all BBF associated peaks of 
$\alpha_{s}(j_1,j_2)> \langle \alpha_{n}\rangle$ and $c^{'}_{fsn}>1$ show a similar feature (not shown), 
that is in average, $\alpha_{s}(j_1,j_2) = 2.5 \pm 0.5 \sim \alpha_{s}(j_3,j_4)\sim 2.5$.
Therefore, after introducing corrections, the BBF-associated small-scale 
and large-scale scaling indices cannot be discriminated. 
We note that the peaks at $\sim$ 0240 UT from s/c 1 (Figure 5b,c) appear due to the  data gaps.

\subsection{Statistical evaluation of BBF and non-BBF associated multi-scale  magnetic fluctuations}

During non-BBF periods $\alpha_{s}(j_3,j_4)$ drops to the noise level
when s/c 1,3 are in the lobe ($B_X\sim 25 $nT) or fluctuates between 1 and 2.5 when s/c 1,3 are in 
the plasma sheet (Figure 5a,b). 
Among others, the time evolution of the scaling indices can be influenced by the location
of the Cluster spacecraft, the duration and the magnitude of perpendicular flows and by the motion 
of the plasma sheet (e.g., thickening or thinning). Despite the complexity, we can separate BBF and
non-BBF associated magnetic fluctuations, based on two different conditions. 

Condition 1 is
based only on small-scale scaling parameters having for BBF-associated fluctuations on both s/c 1,3,
$\alpha_{sn}(j_1,j_2) > \  \langle \alpha_{n}\rangle + 0.3$ and  $c^{'}_{fsn} > 2$  and non-BBF-associated 
fluctuations  $\alpha_{sn}(j_1,j_2) <  \langle \alpha_{n}\rangle + 0.1$ and  $c^{'}_{fsn} < 1.1$. 
It simply means that we relate the observed small-scale magnetic fluctuation-dissipation activity 
to the large-scale magnetic fluctuations. We suppose that on the basis of scaling characteristics
and power of small-scale magnetic fluctuations we can identify BBF and non-BBF associated intervals
without actual examination of velocity data. For testing this supposition we use 
only the parameters  computed from the magnetic field $B_X$ component  measured by s/c 1,3 (Figure 5a-c).
So we select only those values of $\alpha_{s}(j_3,j_4)$  from the whole period (Figure 5b)  
which meet the above conditions for $\alpha_{sn}(j_1,j_2)$ and  $c^{'}_{fsn}$.
The conditional statistics  performed on the large-scale scaling index gives for both s/c 1,3 
$\alpha_{s,BBF}(j_3,j_4) = 2.57\pm 0.05$ for BBF-associated 
and $\alpha_{s,non-BBF}(j_3,j_4) = 1.6\pm 0.4$ for non-BBF associated case.  
 
Now we consider also the velocity data.
Condition 2 is based on thresholding of perpendicular flows, taking
$V_{\perp,xy} \equiv \sqrt {V^{2}_{\perp,x}+V^{2}_{\perp,y}} > 300$ km/s for rapid flows and
$V_{\perp,xy} < 100 $ km/s for non-BBF associated remnant flows. 
Since  the velocity data  is available with a time resolution of $\sim4$s and the time step between
neighbouring sliding windows is $S=4$s, after adjusting times, $\alpha_{s}(j_3,j_4)$ can be compared
to $V_{\perp,xy}$.
It gives
$\alpha_{s,BBF}(j_3,j_4) = 2.6\pm 0.1$  and $\alpha_{s,non-BBF}(j_3,j_4) = 1.7\pm 0.4$. 
All the estimates represent mean values  with  standard deviations. 

Both conditions give the same statistical results on large-scale scaling index. 
From the conditions and the corresponding values of large-scale indices follows 
that, relative to non-BBF associated intervals,  
BBF-associated intervals exhibit  both large-scale and small-scale
spectral steepening and an unambiguous  increase of small-scale power  of magnetic fluctuations.
Therefore we propose 
that small-scale magnetic fluctuations are energized by BBF's. Then the spectral transfer of energy to small scales
is transient, which makes  the position of the spectral break at the expected turnover frequency $f_{c2}$
uncertain. 
Our results suggest that whenever BBF-associated magnetic fluctuations occur, the break at $\sim f_{c2}$
disappears when the small-scale corrections are introduced.  
Otherwise the energy flow to small scales ends at $f_{c2}$, and in this case, at frequencies
$f>f_{c2}$, the measured scaling parameters belong to the  noise. The examination of the LD's at various 
ranges of scales shows that $f_{c2}$ may correspond 
to the approximate Fourier frequencies of $1-3$ Hz (between octaves $j_2$ and $j_3$).

The same analysis of the $B_Y$ and $B_Z$ magnetic field components shows that the observed scaling 
features remain  similar to the $B_X$ case. The small-scale  $B_Y$ and  $B_Z$
fluctuations are also associated mainly with  $V_{\perp,xy}$. 
After introducing corrections as above,
$\alpha_{s}(j_1,j_2) \sim \alpha_{s}(j_3,j_4) \sim 2.6$ for BBF and $\sim 1.7$ for remnant flows, 
the  errors in the same range as above, in conditional statistics of $B_X$. A dissimilarity
appears in $B_Z$  small-scale fluctuations (Figure 7). The small-scale power $c^{'}_{fsn}$
of $B_Z$ fluctuations (Figure 7c) during BBF-associated intervals is several times larger than that of
$B_X$ (Figure 5c) or $B_Y$ (not shown).  Usually $c^{'}_{fsn}(B_Z)>>c^{'}_{fsn}(B_Y)>c^{'}_{fsn}(B_X)$
over time scales less than 1 s.
This is a signature of anisotropy and we will examine it deeper. 

\subsection{Anisotropy of magnetic fluctuations}
The small-scale powers of the magnetic field components 
from s/c 1,3, computed at $j=4$ ($\sim 0.33$ s),
are compared in log-log scatterplots in Figure 8. The comparison of $B_X$ and $B_Y$ components
shows a  balanced scatter around the line $c^{'}_{fsn}(B_X)=c^{'}_{fsn}(B_Y)$ (Figure 8a), though
the increased population of points under that line between $2<c^{'}_{fsn}(B_Y)<10$, indicates
a weak anisotropy.
The scatterplot of $c^{'}_{fsn}(B_X)$ and $c^{'}_{fsn}(B_Z)$ shows strong anisotropy (Figure 8b)
for values $c^{'}_{fsn} > 1.5$.  We also calculated the  scatterplots at  scale $j=6$ ($\sim 1.3$ s)
(Figures 9 a, b), which, after comparison with Figure 8b, indicate scale dependent anisotropy. The
anisotropy increases towards small scales. Since $c^{'}_{fsn}(B_Z)\gg c^{'}_{fsn}(B_X)$
and  $c^{'}_{fsn}(B_Y)\ge c^{'}_{fsn}(B_X)$ in Figure 8, the
small-scale magnetic fluctuations are generated  due to multi-scale transfer of energy
preferentially in the GSM $Z$ or $Z-Y$ direction. The powers at the scale $j=6$ show more isotropy,
even if the scatterplot of $c^{'}_{fsn}(B_X)$ versus $c^{'}_{fsn}(B_Y)$ (Figure 9a) 
indicates that, with increasing
scales the field-aligned power of fluctuations can also increase.

The points depicted by '+' signs in Figures 8b and 9b 
belong to the time period between 0400 and 0410 UT. During this interval a magnetic reconnection event
in the central plasma sheet at 0401 UT was identified by  {\it Baker et al.} [2002]. 
The corresponding '+' points in Figures 8b and 9b, therefore, are associated with the magnetic 
reconnection or/and the tailward flow, leading to a strong anisotropy which  does not 
exhibit the same increase towards the small scales as it is observed for the other points.
Therefore the 
underlying mechanism which drives these fluctuations should be different from the previous scale
dependent one. Since only this reconnection-related short interval shows different anisotropy
characteristics, we cannot say more about the nature of the underlying physical processes. Answering
the question about reconnection or tailward flow associated anisotropy characteristics certainly requires
a thorough statistical analysis. We can speculate, however,
that the observed scale-independent anisotropy features might be related to the specific current systems
or magnetic field topology near the reconnection site [{\it Nagai et al.}, 2003; {\it Runov et al.}, 2003]. 

The small-scale $\alpha_{sn}(j_1,j_2)$ for individual components
are compared in scatterplots in Figure 10 a,b. Again, $\alpha_{sn}(j_1,j_2, B_Z)$ seems to be different 
from $\alpha_{sn}(j_1,j_2, B_X)$ and $\alpha_{sn}(j_1,j_2, B_Y)$. This anisotropy can weaken, however,
because  non-corrected values are depicted and the corrections are larger for $B_X, B_Y$ than for $B_Z$
(Figures 5c and 7c). 

The anisotropy can also be  considered in the coordinate system of the local mean field.
We computed the mean total magnetic field vector  through low-pass
filtering the GSM magnetic field components 
and using standard vector operations transformed the
burst mode magnetic data into mean-field 
perpendicular ($B_{\perp}$) and mean-field aligned ($B_{\parallel}$) components. 
Figures 11a, b show $B_{\perp}$ and $B_{\parallel}$, respectively, computed for averaged
$\langle B\rangle$ at the time-scale $1.3$ s. Figures 11 c, d show the perpendicular 
$c_{fsn}^{'}(B_{\perp})$ and the parallel $c_{fsn}^{'}(B_{\parallel})$ powers, computed at
the scales $j=3$ ($\sim 0.33$ s) and $j=6$ ($\sim 1.3$ s). As above for the GSM $X, Y, Z$
components (Figures 8-10), 
the anisotropy is scale dependent and the small-scale power preferentially develops 
in directions perpendicular to the mean magnetic field. Again, the short interval after $0400$
UT shows a different kind of reconnection or tailward flow related behavior. The large-scale
($j=6$) perpendicular power remains larger than the parallel power only during this interval.
The similarity of the statistical results obtained in GSM and mean field coordinate systems 
can be explained as follows: for open, stretched magnetic field lines 
$B_{\parallel}$ resembles $B_X$ and $B_{\perp}$ is similar to $\sqrt{B_{Z}^2+B_{Y}^2}$ (where
$B_{Y}$ and $B_{Z}$ are the GSM components of the magnetic field). 
Dissimilarities can appear  close to the neutral sheet when $B_X\to0$. 

\subsection{Two-point conditional statistics}

Let us investigate now the effect of the transitory character of the fluctuations or their drivers. 
We compare  magnetic fluctuations observed at the same time on two spatially separated Cluster spacecraft.
Figure 12 compares  the parallel $\alpha_{s\parallel}(j_3,j_4)$ 
and the perpendicular  
$\alpha_{s\perp}(j_3,j_4)$ scaling indices from s/c 1 and 3. 
The scatterplot in (Figure 12a) involves all the magnetic data 
from s/c 1,3 and forms an irregular cloud of points. We applied two-point conditional statistics
to find BBF- and non-BBF-associated points in that cloud. We used the same small-scale condition 1  
combined with the velocity condition 2 as above (both should give the same results), and
required the fulfilment jointly at s/c 1,3  at the same time. In consequence of the joint conditional 
requirement the amount of data points in (Figures 12 b,c) is significantly reduced. It can be explained
partially by  a relatively high level of fluctuations of the considered parameters. For example, the 
perpendicular velocity wildly fluctuates during the rapid flows (e.g. Figure 5d)
which makes  the comparison with the small-scale parameters more difficult. The omission of the 
velocity condition, however, does not add many more points to the Figures 12 b,c. Therefore, we suppose
that fewer joint data points are found by two-point conditional statistics 
because the two spacecraft are often in physically different regions,
exhibiting different scaling features. It indicates that the magnetic fluctuations and their sources
are transient and limited in space. 

The statistical evaluation of the  BBF-associated data points
in Figure 12b gives averages $\alpha_{s\perp}(j_3,j_4) = 2.4 \pm 0.3$ and 
$\alpha_{s\parallel}(j_3,j_4) = 2.7 \pm 0.3$ which is consistent with the previous estimations using the
components $B_{X,Y,Z}$ (dashed lines).  The '+' signs in Figure 12b represent measurements during 
the reconnection event, discussed above. 

Figure 12c shows the  non-BBF-associated data points. Here the '+' signs represent measurements near the plasma
sheet boundary layer or in the lobe ($B_X \ge 20$ nT) while the remaining points  are inside of the plasma
sheet ($B_X \le 15$ nT). The statistical evaluation of the latter gives
$\alpha_{s\perp}(j_3,j_4) = 1.7 \pm 0.3$ and 
$\alpha_{s\parallel}(j_3,j_4) = 1.7 \pm 0.4$, which is again consistent with the previous estimations 
using the components $B_{X,Y,Z}$ (dashed lines). 
   
Let us summarize the main results at the end of this section.
We used  a wavelet method to study the small-scale properties of magnetic fluctuations. We found that
the energy transfer to the smallest available time scales is associated with BBFs. The scaling properties
of those fluctuations are recoverable from burst mode magnetic fluctuations by using an embedding technique.
Small-scale  fluctuations show an increased anisotropy relative to the mean magnetic field. 
The anisotropy is observed mainly in the normalized power of the magnetic fluctuations, while it is less
evident in the spectral scaling index. The transient character of the physical processes or 
driving mechanisms makes  a robust estimation of the basic characteristics of magnetic turbulence difficult
in the plasma sheet.

\section{Discussion}
A complete characterization of MHD turbulence requires not only a statistical description of the velocity
field, but also a statistical study of other measurable quantities, e.g. magnetic
field, density or  temperature fluctuations. In this paper we analysed a ~4 hour long period of
high quality burst mode (67 Hz) magnetic field data, which allowed an immersion to the scale or frequency 
ranges which are not available in velocity measurements. We considered time scales from $\sim 0.08$ to 
$5$ s, which, assuming $500$ km/s velocities during rapid flows,  correspond roughly to spatial 
scales of $40 - 2500$ km in the flow direction. 
The average velocities are smaller than $500$ km/s, so the largest scale
can be significantly smaller than $2500$ km. Also, both the velocities and the estimated spatial scales
are smaller in other than the maximum velocity direction or during non-BBF intervals. 
Since the average separation of the Cluster spacecraft was $\sim 1500$ km
during 2001,  we used principally a single-point technique to analyse scaling characteristics of
magnetic fluctuations from s/c 1,3. The usual multi-point statistical approach, commonly
available in laboratory measurements, is not available  in the plasma sheet at those scales and 
the conversion of the scaling characteristics of magnetic fluctuations in time into scaling laws 
in wavenumber space is  possible  only through additional hypotheses. The Taylor hypothesis assumes that 
the conversion is possible if the spatial fluctuations on a given scale pass over the spacecraft 
faster than they typically evolve in time. In the plasma sheet this can be the case during fast
BBF's [{\it Horbury}, 2000] or during strong global bulk flows which may occur after a dynamical reconfiguration
of the magnetotail [{\it Borovsky and Funsten}, 2003]. Otherwise, the random velocities of the fluctuations
advect spatial structures over the spacecraft.
A more realistic approach is the so-called random
sweeping model (RSM), which takes  the basically random velocities of the flow into account and allows
to obtain  statistical information on spatial structures. 

{\it Borovsky et al.} [1997] have used RSM and found that the spectral index, describing the scaling properties
of magnetic fluctuations in the plasma sheet at the frequency range of $1/4 - 1/150$ $ s^{-1}$, is the
same in both frequency or wavenumber representations, that is, $k^{-\alpha} \sim f^{-\alpha}$.
The 'large-scale' ($\sim 1/0.3 - 1/5$ $ s^{-1}$) used in this paper partially overlaps with the scales
of {\it Borovsky et al.} [1997] paper. Note that, RSM is supposed to work when the large scale velocity field
drives the small-scale fluctuations. This seems to be valid in our case due to the high correlation between
BBFs and small-scale magnetic fluctuations. Therefore we suppose, that at least at the scales ($j_3, j_4$)
the equality $k^{-\alpha} \sim f^{-\alpha}$ holds also in our case and the estimated values of
$\alpha$ can be compared with the power-law dependence of power spectra in MHD turbulence. We have
to underline, however, that RSM is only a model  and its validity should be proven comparing real
spatial and time structures. In absence of such comparisons the interpretation of scaling
indices in terms of MHD turbulence remains disputable. At the same time, we consider the observed
anisotropy effects to be more reliable. Let us start the comparison of observations and theory, therefore,
with the evaluation of the influence of a local mean magnetic field on anisotropy of fluctuations.

The clear signs of the interaction of rapid flows with the magnetic field in the plasma sheet are
the BBF-associated dipolarization of field lines (increase of $B_Z$) and  the breaking and diversion 
of the flows around the dipolar inner magnetosphere. The presence of a mean magnetic field
in the plasma sheet cannot be eliminated by coordinate transformations. This is in contrast with
hydrodynamic turbulence, where the anisotropy usually ceases in the coordinate system of the 
average flow [{\it Biskamp}, 2003]. The classical Kolmogorov theory for hydrodynamic and the Iroshnikov-Kraichnan
theory for MHD turbulence assume isotropy of the inertial-range energy cascade in Fourier space,
exhibiting scalings $E(k) \sim k^{-5/3}$ and $\sim k^{-3/2}$, respectively ($E(k)$ is the amount
of energy between wavenumbers $k$ and $k+dk$ divided by $dk$). 

{\it Shebalin et al.} [1983] studied incompressible MHD anisotropies arising in  wave vector space 
in the presence of a mean magnetic field. 
They studied the interaction of opposite-travelling wave packets and found that, in 
wave vector space, those interactions produce modes with wavevectors preferentially perpendicular to 
the mean magnetic field. 
{\it Goldreich and Sridhar} [1995] proposed a balance condition between parallel and perpendicular modes
(parallel propagating
Alfv\'{e}n waves and perpendicular eddy motions). On this basis it was shown that 
a scale dependent anisotropy appears, namely
$k_{\parallel} \sim k_{\perp}^{2/3}$ (or $l_{\parallel} \sim l_{\perp}^{2/3}$, where $l\sim1/k$ is the 
characteristic scale of eddies in parallel and perpendicular directions), 
which means that the anisotropy is increasing with decreasing
scale. Moreover, it was shown that the perpendicular fluctuations exhibit  Kolmogorov scaling 
$E(k) \sim k^{-5/3}$. In other words, the mean magnetic field does not influence the perpendicular
hydrodynamic cascade formed by eddies. The coupled parallel fluctuations are essentially produced by
waves. In the case of compressible MHD the Alfv\'{e}n modes follow the anisotropic Goldreich-Sridhar
scaling while fast modes exhibit isotropy [{\it Cho et al.}, 2003]. 

The scale dependent anisotropy proposed by {\it Goldreich and Sridhar} [1995]  would represent a
way   the relative importance of Alfv\'{e}n waves and eddy motions might be recognized. {\it Matthaeus et
al.} [1998] tested this assumption  using Fourier spectral methods 
for  MHD equations in a periodic cube. They have found,  that for incompressible,
weakly compressible and driven MHD turbulence, anisotropy scales
linearly with the ratio of fluctuating to total magnetic field strength. 
Their results therefore seem to be inconsistent  
with the Goldreich-Sridhar model. {\it Cho and Vishniac} [2000] argued, however, that the Fourier technique used
by {\it Matthaeus et el.} [1998] smooths out the true scaling relation for anisotropy. 
It happens because the anisotropic eddies or wave packets are elongated along magnetic field lines. When
they  follow the large-scale variation of the magnetic field, they point in different directions.
Consequently, the information about the eddy shapes is lost when a global transformation such as the
Fourier transformation is applied.
{\it Cho and Vishniac} [2000]
showed that, when the eddy shapes relative to the local magnetic field are analysed in real space, the
Goldreich-Sridhar scaling is recovered.  

For what follows, it is important that not only the velocity statistics exhibits the Goldreich-Sridhar
anisotropy, but also the magnetic field statistics [{\it Cho and Lazarian}, 2004]. 
If we suppose similarity between frequency and wavenumber scalings, then 
our results suggest that the observed magnetic turbulence in the plasma sheet
also resembles the anisotropy of Goldreich-Sridhar model. The small-scale magnetic turbulence is 
stronger in perpendicular directions. 
It is mainly visible when the relative powers of fluctuations are compared  (Figures 8 and 11). 
Then, except for the short interval when magnetic reconnection and tailward flow occurs, 
the anisotropy is scale dependent (Figures 8,9). The applied wavelet technique
ensures that the scale-dependent anisotropy is not smoothed out. Namely, over each considered scale range
the local trend (local mean field) is cancelled by the analysing wavelet having 
a proper number of vanishing moments. 
The observed anisotropy seems to indicate that the small-scale eddy turbulence is stronger than 
the Alfv\'{e}nic turbulence.	
This is in agreement with the results of {\it Borovsky and Funsten} [2003], who claimed that turbulence of eddies
rather than the turbulence of Alfv\'{e}n waves prevails in the plasma sheet. Their observation is based
on three different arguments: (a.) Alfv\'{e}n wave turbulence produces scale sizes that are larger
than the vertical size of the plasma sheet, whereas eddy turbulence does not; (b.) the autocorrelation
length of the turbulence is typical for eddy turbulence; (c.) the low Alfv\'{e}n ratio (ratio of the
kinetic energy of the flow to the energy density of magnetic fluctuations) indicates 2D eddy structures.
All these results might yield a picture of  the Goldreich-Sridhar model and quasi-2D MHD turbulence. 
The quasi-2D MHD picture is supported by 
numerical studies of high Reynolds number  turbulence which predict that the spatial distribution
of the turbulent fields is more intermittent in 2D than in 3D MHD ([{\it Biskamp}, 2003], and references
therein). Since BBF-associated magnetic fluctuations are stretched down to the small scales $(j_1,j_2)$
(Figures 5,7), where the anisotropy becomes stronger and turbulence might be essentially  2D, 
an increase of the small scale intermittence can  be expected, too. Indeed, 
{\it V\"{o}r\"{o}s et al.} [2003] have
shown that small scale magnetic field intermittence increases during BBF-associated intervals. 
Here, some more  words of caution have to be added.
First of all, the  importance of multi-scale dynamics in turbulence has to be underlined. Since
the Goldreich-Sridhar anisotropy is scale dependent, at sufficiently large scales this kind of anisotropy
can disappear, and even Alfv\'{e}nic turbulence can become stronger. Figures 9a and 11 d
show that the large scale power of parallel ($B_{X}$ or $B_{\parallel}$ ) fluctuations  
can be equal with or slightly overpower the perpendicular components ($B_{Y},B_{Z}$ or $B_{\perp}$).
The development of 
anisotropy and 2D structures can also be dependent on  the actual Reynolds numbers,  helicity, Alfv\'{e}n ratio 
[{\it Oughton et al.}, 1994] or be influenced by the presence of transverse pressure-balanced
magnetic structures or transverse velocity shears [{\it Ghosh et al.}, 1998].   
In this respect it is important to note that in our case the mean field direction is constantly changing
due to the time-varying  contribution of magnetic field components to the total magnetic field 
(compare Figures 5,7). Nevertheless, the small scale power of the $B_Z$ (vertical) fluctuations is the
strongest all the time. The vertical GSM direction, however, 
is not always the perpendicular direction to the
mean magnetic field. Therefore the effect of the mean field, even if it represents a robust way
for the development  of scale-dependent anisotropy 
has to be complemented by other mechanisms, which can produce strong fluctuations in perpendicular 
directions. For example the above mentioned transverse magnetic structures or velocity shears can also 
redirect the initially parallel-propagating Alfv\'{e}n waves to highly oblique waves 
[{\it Ghosh et al.}, 1998]. In the plasma sheet, BBF-associated dipolarization, increase of $B_{Z}$
(or $B_{Y}$ when the current sheet is tilted), can represent such a transverse magnetic structure.
Also, the relative thinness of the plasma sheet in the z direction and the associated boundary effects
can produce anisotropy statistics in plasma sheet turbulence [{\it V\"{o}r\"{o}s et al.}, 2004].   
To evaluate the possible contributions of these mechanisms a wider statistical study is required, however.
 
Let us consider now the  estimated scaling indices $\alpha$ in comparison with the predictions of
Goldreich-Sridhar model. When the anisotropy effects are neglected,  
the conditional statistics explained above gives for  BBF-associated flows the large scale index
$\alpha_{s}(j_3,j_4) = 2.6 \pm 0.1$, independent on magnetic field component. 
It is definitely different from
the Goldreich-Sridhar model which predicts Kolmogorov scaling with $\alpha \sim 5/3$ in perpendicular direction. 
We could advance
an argument that the Goldreich-Sridhar anisotropy is weaker at the scales $(j_3,j_4)$, but the
small-scale $\alpha_{s}(j_1,j_2)$ reaches the same range of values when the estimated $c^{'}_{fsn}$ is
large enough and allows to introduce corrections. $\alpha_{s}(j_3,j_4)$ decreases and approaches
the Kolmogorov scaling during non-flow or post-BBF intervals, when the ratio of the large-scale mean field
to the magnetic fluctuations becomes larger and the conditions for the development of Goldreich-Sridhar
anisotropy are more favourable. In this case the estimated $\alpha_{s}(j_3,j_4)=1.7 \pm 0.4$. The large
error appears because the post-flow  intervals usually correspond to transient, probably decaying
turbulence. This also causes that one cannot distinguish between the Kolmogorov and Iroshnikov-Kraichnan models.
The high value of $\alpha_{s}(j_3,j_4)$ during BBF-associated intervals does not necessarily mean that
the Goldreich-Sridhar phenomenology fails. Magnetic fluctuations are subject to an additional constraint
which acts during rapid flows only. The limited vertical dimension of the flow channel 
or the plasma sheet thickness can modify
the actual value of $\alpha$ increasing it to $\alpha \sim 3$, which is a limiting value for fully 
developed 2D hydrodynamic turbulence [{\it Volwerk et al.}, 2003, 2004, {\it V\"{o}r\"{o}s et al.}, 2004]. 

\section{Conclusions}
The work described here demonstrates the ability of the windowed wavelet estimator for obtaining reliable
estimates of the effective second-order scaling parameters $(c_{f}, \alpha)$ even in situations when the noise level 
is relatively high. 
Magnetic turbulence seems to be steady only during relatively short time intervals (a few minutes), otherwise
$\alpha $  exhibits nonstationary behaviour at all scales, which can partially explain the large
scatter of previously estimated values of scaling indices. As we observed magnetic turbulence associated
with bursty flows and post-flow intervals, 
our findings agree well with previous knowledge that the flow of plasma in the
Earth's magnetotail is intrinsically unsteady [{\it e.g. Baumjohann}, 2002]. Joint conditional
statistics from two Cluster spacecraft s/c 1,3 shows a high level of spatial intermittence
which is related to the differing scaling features in physically different regions in the plasma sheet.

The examination of BBF-associated high frequency magnetic fluctuations within a sliding window
showed that  the small-scale 
($j_1, j_2$) and large-scale ($j_3, j_4$) scaling indices
are equal within error estimations. {\it Volwerk et al.} [2003, 2004] have obtained similar scaling exponents for the 
time scales 1-13 s. It means that the spectral break predicted by {\it Milovanov et al.} [2001] 
for the scales $< 1/f_{c2} \sim 10$ s is absent and, 
using the notation from the Introduction, $\alpha_2 \sim \alpha_3$.
During non-BBF periods the small-scale $\alpha_3 \sim \alpha(j_1, j_2)$ 
aproaches the noise level of the magnetometer, 
which introduces an artificial break at the scale $\sim 1$ s. Obviously, it cannot be interpreted 
in terms of 
a new scaling regime describing a scale dependent physics. 

We found that, for the scales analysed, the conversion of the observed scalings from frequency to wavenumber
representation is model dependent and cannot be verified by direct comparison of spatial and
temporal variations. It makes the interpretation of the results in terms of MHD phenomenology more
difficult. The  anisotropy features of magnetic fluctuations provided additional tools to
facilitate the interpretation of the results. The anisotropy is better visible in the normalized 
power of fluctuations $c_f$ than in the scaling index $\alpha$.
It was shown that large-scale magnetic fluctuations are more isotropic while 
small-scale magnetic turbulence evolves in  direction perpendicular to the mean magnetic field, 
usually in the GSM $Z$ or $Z-Y$ 
direction, and the observed anisotropy is scale dependent. 
It partly agrees  
with the predictions of Goldreich-Sridhar  model of MHD turbulence, but the additional effect of transverse
magnetic structures or velocity shears cannot be excluded. On this basis we think the
turbulence in the plasma sheet is a mixture of Alfv\'{e}nic wavy turbulence and of eddy 2D turbulence. 
The latter becomes stronger and more intermittent with decreasing scales. 

The occurrence of  reconnection and/or  rapid tailward flow
can increase the level of anisotropy. The build-up of a flow channel seems to modify the 
actual value of the scaling indices, indicating the presence of additional constraints on Alfv\'{e}nic and/or
eddy turbulence. These are observations which should be verified on larger data sets.

{\it Acknowledgement} \\
The authors acknowledge the use of the code for the estimation of scaling exponents developed
by P. Abry and his colleagues.

%
%

\newpage

%
%
%


%
%
%
\pagebreak
\begin{figure}[tb]
\centerline{
\includegraphics[width=3.2in]{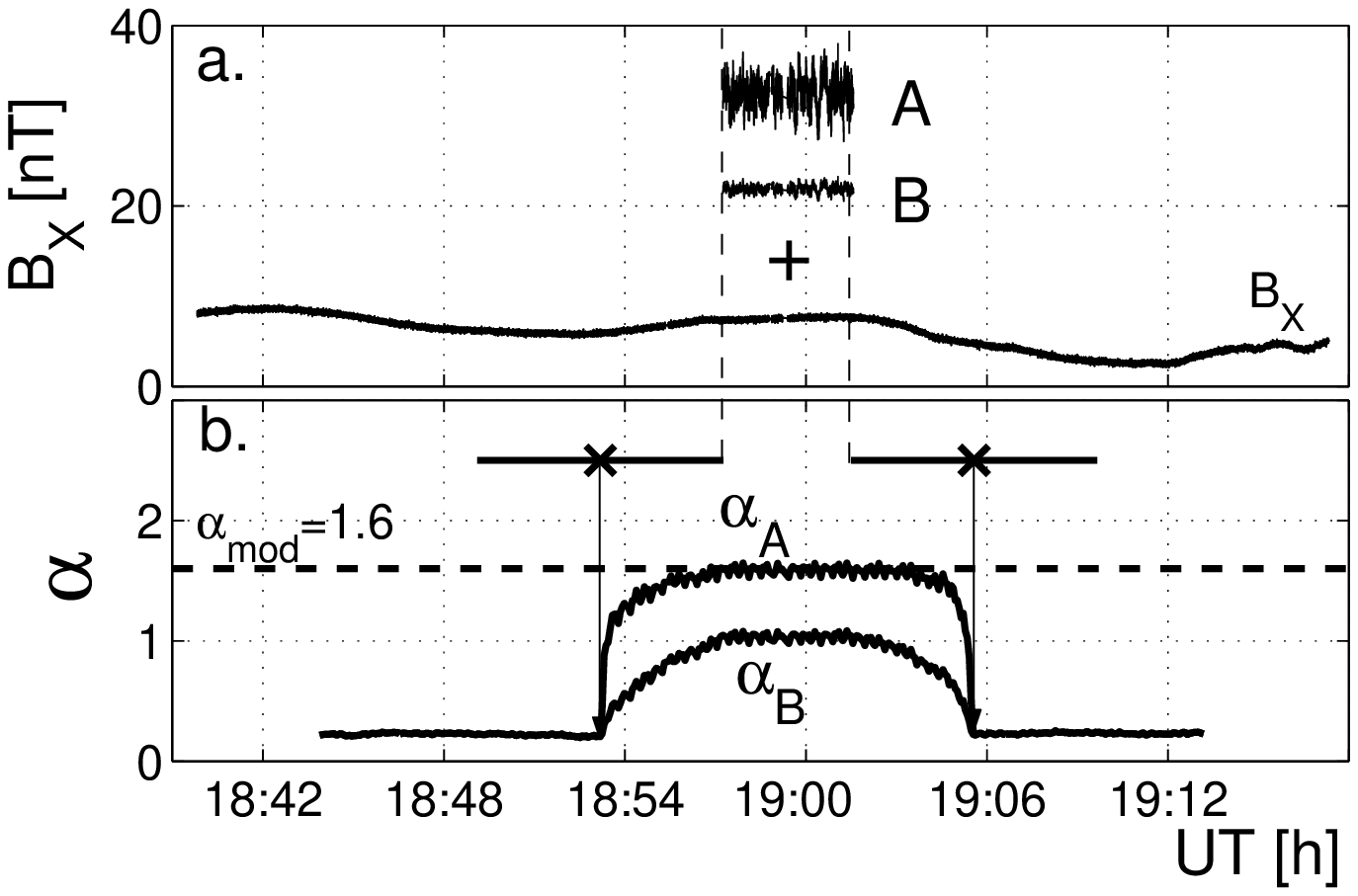}
}
\caption{Estimation of the scaling index $\alpha$; (a) Fractal signals 
A and B to be embedded and quiet time $B_X$ measured by Cluster 3 
on September 7, 2001; the generated fractal signals  A and B have 
the same $\alpha_{mod}=1.6$, but different amplitudes,  
(b) $\alpha_A$ and $\alpha_B$ from mixed signal; (the dashed line corresponds to the
theoretical value $\alpha_{mod}=1.6$)}
\end{figure}

\pagebreak
\begin{figure}[tb]
\centerline{
\includegraphics[width=3.2in]{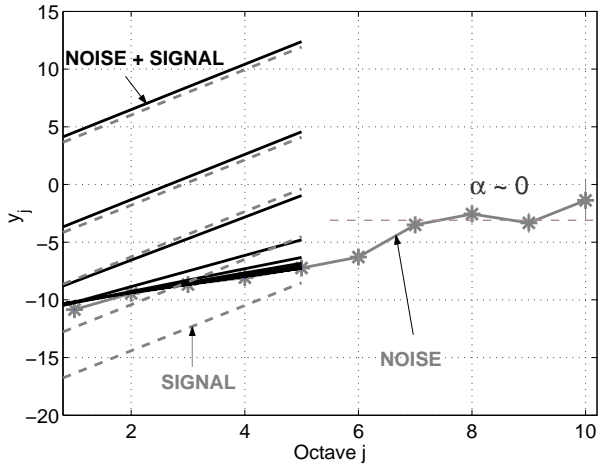}
}
\caption{The logscale diagram; The scaling properties of signal, noise and signal+noise 
cases are shown when the signal to noise ratio is changed; $y_j\equiv log_2 \mu_j$ (Eq.1)}
\end{figure}

\pagebreak
\begin{figure}[tb]
\centerline{
\includegraphics[width=3.2in]{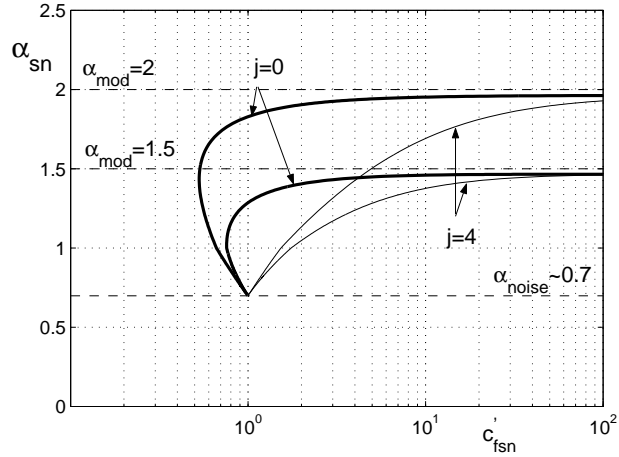}
}
\caption{The $\alpha_{sn}(c^{'}_{fsn})$ relationship (correction curves) 
for model fractal signals with $\alpha_{mod}=1.5$ and $2$, computed
at the octaves $j=0$ and $j=4$}
\end{figure}


\pagebreak
\begin{figure}[tb]
\centerline{
\includegraphics[width=3.2in]{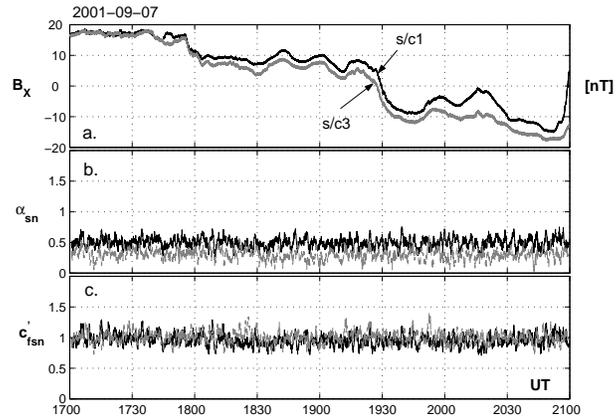}
}
\caption{Estimation of the quiet time noise level of the scaling parameters; 
a.) $B_X$ component of the magnetic field on  September 7, 2001
b.) $\alpha_{sn}$ 
c.) $c^{'}_{fsn}$; all for s/c 1 and 3 }
\end{figure}


\pagebreak
\begin{figure}[tb]
\centerline{
\includegraphics[width=3.2in]{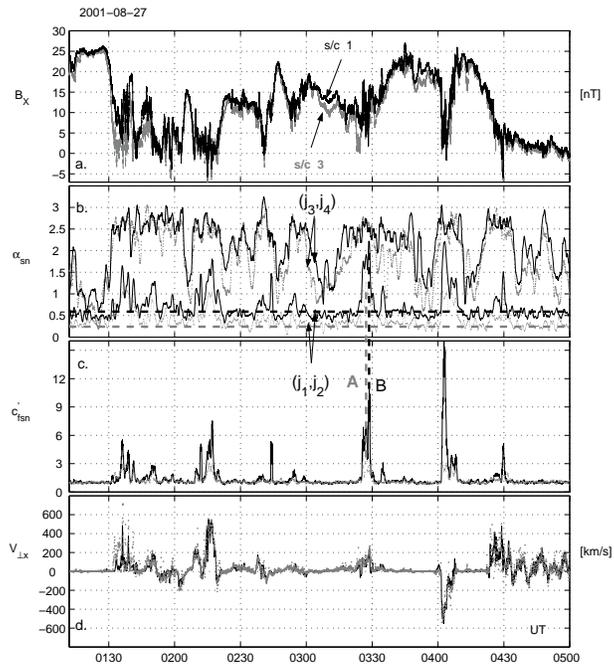}
}
\caption{Two-point (s/c 1,3) estimation of the scaling parameters in comparison with the magnetic field
and velocity measurements; a.) the $B_X$ components of the magnetic field; b.) $\alpha_{sn}$ estimated
at  small scales $(j_{1}, j_{2})\sim (0.08,0.33)$s and at  large scales $(j_{3}, j_{4})=(0.7,5)$s;
c.) $(c^{'}_{fsn})$ estimated at the scale $j=4\sim 0.33$ s; the capital letters A and B show two cases
when corrections to small-scale $\alpha$ are introduced in Figure 6; 
d.) x-component of perpendicular velocity}
\end{figure}

\pagebreak
\begin{figure}[tb]
\centerline{
\includegraphics[width=3.2in]{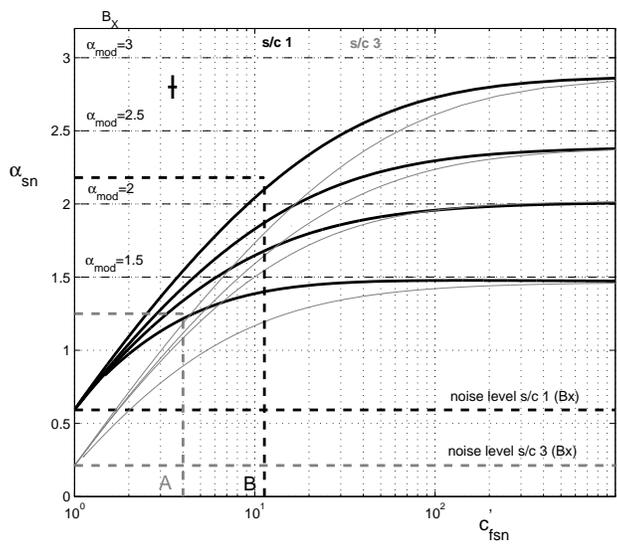}
}
\caption{Correction curves computed for model fractals with $\alpha_{mod} = 1.5, 2, 2.5, 3$;
the correction curves reach asymptotically the theoretical values $\alpha_{mod}$ for 
$c^{'}_{fsn}\to \infty$ and the noise level for $c^{'}_{fsn}\to 1$; the letter A corresponds to s/c 3 while
B to s/c 1;  the error bars at the left-top
corner arise from the uncertainty of parameter estimation in logscale diagram and from the uncertainty
of the magnetometer noise level estimation}
\end{figure}

\pagebreak
\begin{figure}[tb]
\centerline{
\includegraphics[width=3.2in]{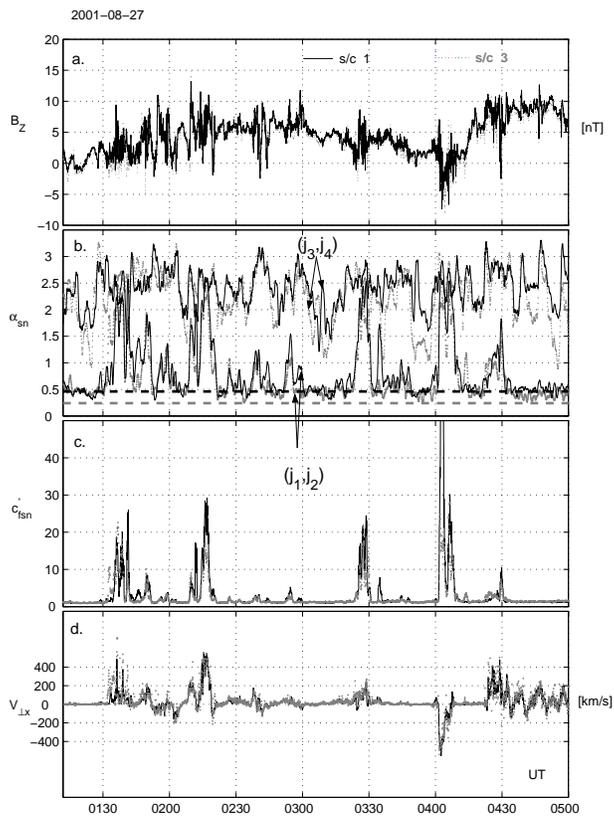}
}
\caption{Same as in Figure 5, but for the $B_Z$ component of magnetic field}
\end{figure}

\pagebreak
\begin{figure}[tb]
\centerline{
\includegraphics[width=3.2in]{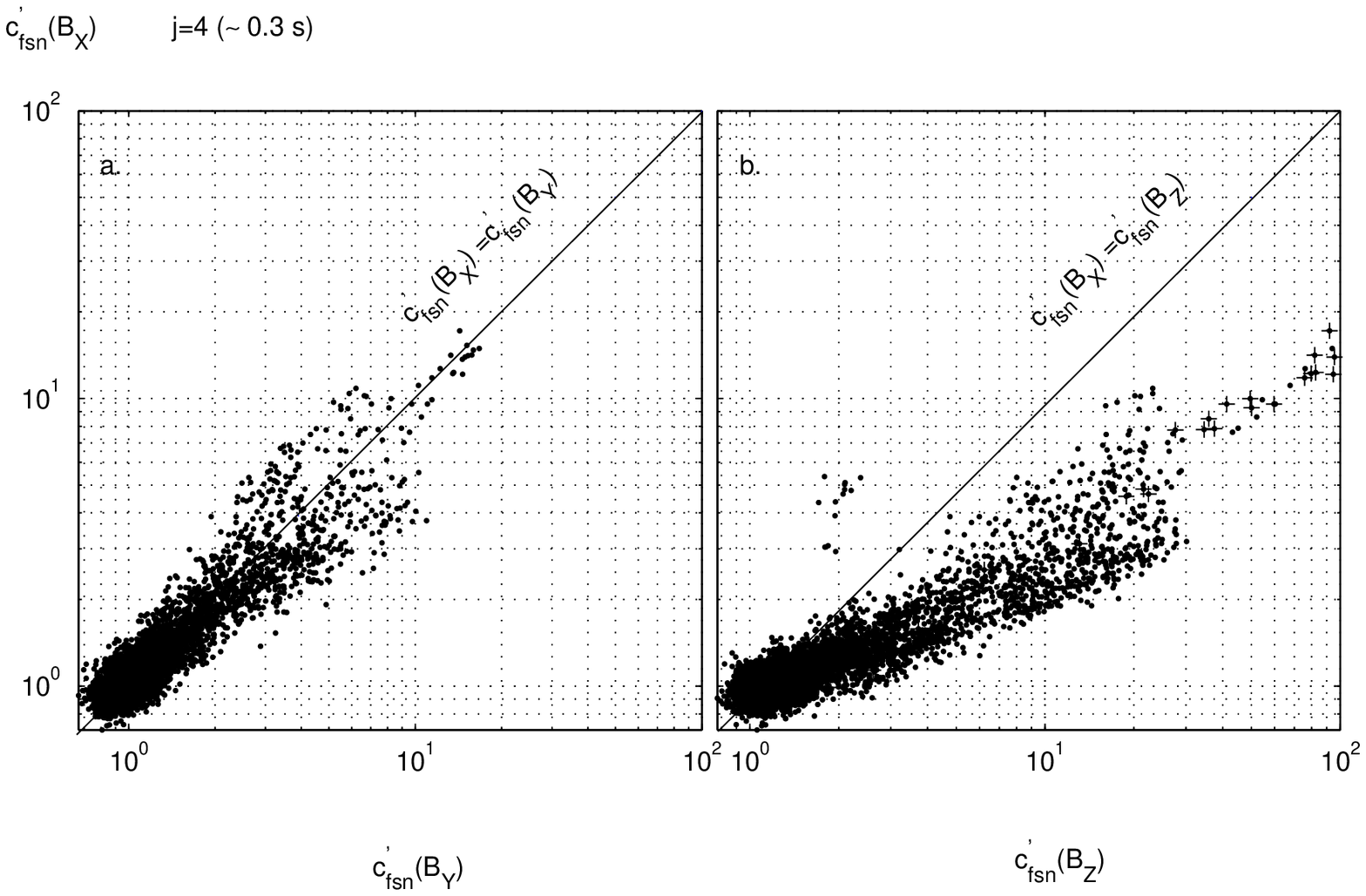}
}
\caption{Log-log scatterplots of powers at the scale $j=4 \sim 0.33$ s; 
a.)$c^{'}_{fsn}(B_X)$ versus $c^{'}_{fsn}(B_Y)$;
b.) $c^{'}_{fsn}(B_X)$ versus $c^{'}_{fsn}(B_Z)$ - the crosses correspond to 
an interval of magnetic reconnection}
\end{figure}

\pagebreak
\begin{figure}[tb]
\centerline{
\includegraphics[width=3.2in]{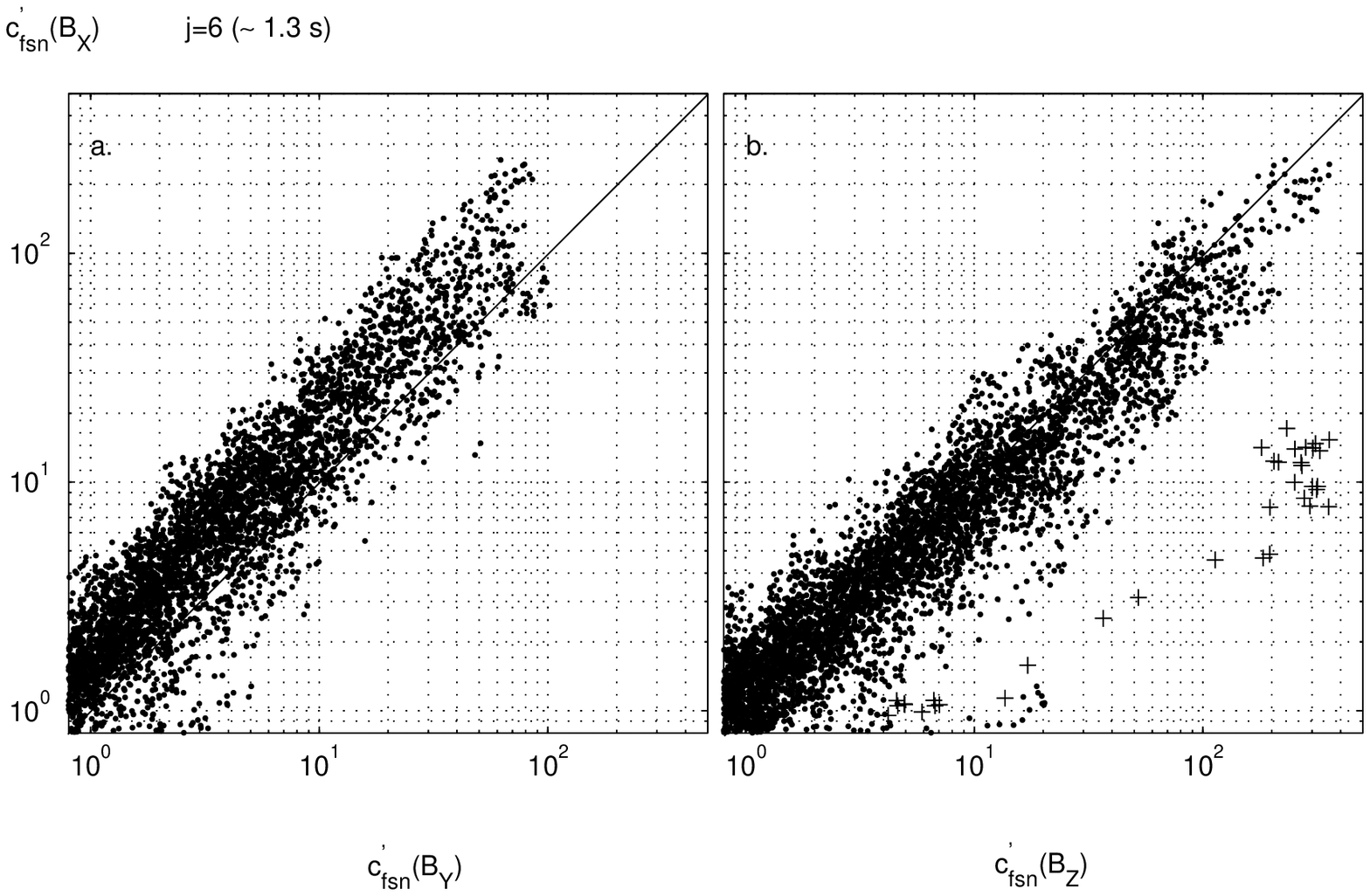}
}
\caption{Log-log scatterplots of powers at the scale $j=6 \sim 1.3$ s; 
a.)$c^{'}_{fsn}(B_X)$ versus $c^{'}_{fsn}(B_Y)$;
b.) $c^{'}_{fsn}(B_X)$ versus $c^{'}_{fsn}(B_Z)$ -
the crosses correspond to an interval of magnetic reconnection}
\end{figure}


\pagebreak
\begin{figure}[tb]
\centerline{
\includegraphics[width=3.2in]{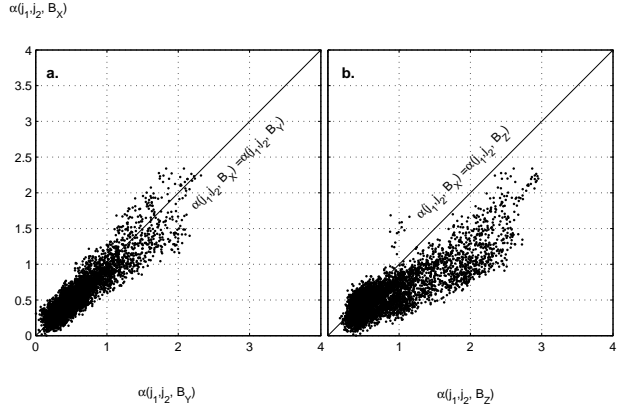}
}
\caption{Small-scale $(j_1,j_2)$ scatterplots of the scaling indices;
a.) $\alpha_{sn}(B_X)$ versus $\alpha_{sn}(B_Y)$;
b.)$\alpha_{sn}(B_X)$ versus $\alpha_{sn}(B_Z)$}
\end{figure}

\pagebreak
\begin{figure}[tb]
\centerline{
\includegraphics[width=3.2in]{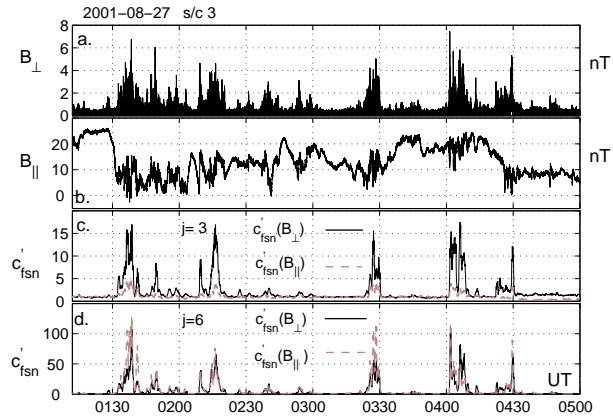}
}
\caption{Magnetic fluctuations in the mean magnetic field coordinate system;
a.) the perpendicular component ($B_{\perp}$);
b.) the parallel component ($B_{\parallel}$);
c.) the normalized powers of $B_{\perp}$ and $B_{\parallel}$ at the time scale $0.33$s;
d.) the normalized powers of $B_{\perp}$ and $B_{\parallel}$ at the time scale $1.3$s;}
\end{figure}


\pagebreak
\begin{figure}[tb]
\centerline{
\includegraphics[width=3.2in]{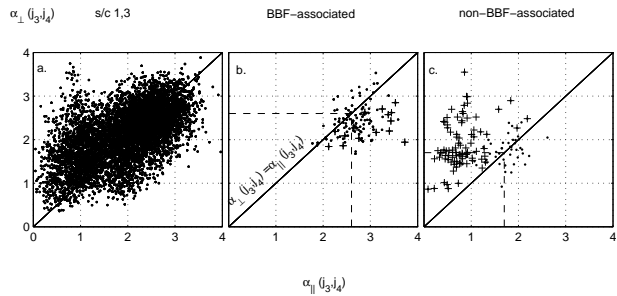}
}
\caption{Joint conditional statistics in a coordinate system parallel and perpendicular to the local mean
magnetic field at scales $(j_3,j_4)$; a.) scatterplot of all available points from s/c 1,3;
b.) scatterplot of BBF-associated data points - the crosses correspond to an interval of 
magnetic reconnection; 
c.) scatterplot of non-BBF-associated data points - the crosses correspond to the intervals when 
$|B_X| > 20$ nT, while the points represents intervals with $|B_X|\le 15$ nT}
\end{figure}


\end{document}